\documentclass[a4]{article}
\usepackage{arxiv}
\usepackage[utf8]{inputenc} 
\usepackage[T1]{fontenc}    
\usepackage{booktabs}       
\usepackage{amsfonts}       
\usepackage{nicefrac}       
\usepackage{microtype}      
\usepackage{textcomp}       

\usepackage{times}
\usepackage{graphicx}
\usepackage{subcaption}                 
\usepackage{amssymb}
\usepackage{url}
\usepackage{hyperref}
\usepackage{geometry}
\usepackage{amsmath}
\usepackage{physics}
\usepackage{verbatim}                   
\usepackage[ruled,vlined]{algorithm2e}  
\usepackage{fancyhdr}
\usepackage{cleveref}   

\graphicspath{{figures/}}

\fancypagestyle{firststyle}
{
   \fancyhf{}
   \chead{}
}

\begin{document}

\title{Implicit Hybrid Quantum-Classical CFD Calculations using the HHL Algorithm}

\author{Leigh Lapworth \\
\\
Rolls-Royce plc \\
Derby, UK \\
\today
\\
\\
leigh.lapworth@rolls-royce.com  \\
}

\maketitle
\thispagestyle{firststyle}     

\begin{abstract}
Implicit methods are attractive for hybrid quantum-classical CFD solvers
as the flow equations are combined into a single coupled matrix that is solved
on the quantum device, leaving only the CFD discretisation and matrix 
assembly on the classical device.
In this paper, an implicit hybrid solver is investigated using
emulated HHL circuits. The hybrid solutions are compared with classical solutions
including full eigen-system decompositions.
A thorough analysis is made of how the number of qubits in the HHL eigenvalue
inversion circuit affect the CFD solver's convergence rates. 
Loss of precision in the minimum and maximum eigenvalues have different
effects and are understood by relating the corresponding eigenvectors to
error waves in the CFD solver. An iterative feed-forward mechanism is identified
that allows loss of precision in the HHL circuit to amplify the associated error waves.
These results will be relevant to early fault tolerant CFD applications
where every (logical) qubit will count.
The importance of good classical estimators for the minimum and maximum
eigenvalues is also relevant to the calculation of condition number
for Quantum Singular Value Transformation 
approaches to matrix inversion.
\end{abstract}

%
\section{Introduction}
\label{sec-intro}

Computational Fluid Dynamics (CFD) is recognised as a crucial enabler
to national productivity and competitiveness \cite{walport2018computational}.
It has also been recognised as a discipline where quantum computing is expected 
to have a major impact although 
{\it "...this is not a short-term endeavor"} \cite{givi2020quantum}.
There is already a body of research investigating quantum CFD algorithms
\cite{steijl2019quantum, kacewicz2006almost, costa2019quantum,
gaitan2020finding, suau2021practical, ljubomir2022quantum, lu2020quantum} using theoretical
fault-tolerant analysis and emulated circuits.
NISQ era variational algorithms also exist, including demonstrations
on physical devices \cite{lubasch2020variational, kyriienko2021solving}.
A characteristic of the previous work is that test cases are
chosen that result in matrices with repeated entries and, often, a Toeplitz structure.
Broader non-CFD quantum linear equation solvers have also focused on matrices with
a small number of real valued degrees of freedom. The 8x8 Toeplitz matrix studied by
\cite{vazquez2020enhancing} needed only two real numbers to define all the non-zero entries.
In previous work \cite{lapworth2022hybrid}, the author considered a real world CFD test case and studied
the Poisson type pressure correction matrix common to many CFD solvers that use
the SIMPLE algorithm and its derivatives \cite{patankar1972calculation, ammour2018subgrid}.
Matrices with up to 288 entries, of which 176 were unique, were analysed using an
emulated HHL algorithm \cite{harrow2009quantum}.

Whilst the SIMPLE based schemes have been hugely successful for many years,
and continue to be so, they exemplify a key characteristic of classical CFD solvers:
direct matrix inversion is intractable for all but the smallest cases.
Rather than seek to migrate today's classical algorithms to a quantum computer,
this paper asks what would a CFD algorithm look like if full matrix inversion were possible,
and efficient, for the largest matrices.
To some extent the answer is already known from classical CFD solvers which use implicit schemes that
require fewer, more computationally expensive, iterations than explicit or segregated schemes.
\cite{misev2018development} showed an average factor of 3 increase in memory for the industrial
CFD code HYDRA \cite{lapworth2004hydra, amirante2021multifidelity}, achieving a speed-up of over 
a factor of 8 relative to the explicit multi-grid code \cite{moinier2002edge}.

Both implicit and explicit codes
use acceleration techniques such as multi-grid \cite{wesseling1982robust} and linear algebra 
techniques such as the Conjugate Gradient (CG) family
\cite{hestenes1952methods, fletcher1976conjugate, van1992bi} to give tight convergence of
their linearised matrix equations. 
Krylov subspace methods such as CG codes are only guaranteed to converge
in infinite precision and may require as many steps as the rank of the matrix.
Hence, industrial CFD codes use preconditioning often with sophisticated parallel
implementations \cite{idomura2018communication}.
Preconditioning is not always capable of ensuring convergence of the CG solver. In such cases 
the GMRES scheme \cite{saad1986gmres} is often adopted which minimises a residual over
a stored set of orthogonal Krylov basis vectors. Unfortunately, this is memory intensive and only
a small number of basis vectors can be stored before GMRES has to be {\it restarted}. 
\cite{misev2018development} stored only 10 basis vectors before restarting.
Many of these techniques are available in massively parallel supercomputing libraries
such as PETsc \cite{balay1998petsc}.

As cited, classical supercomputing has evolved highly efficient 
parallel matrix solvers that obviate 
the need for direct matrix inversion. However, quantum solvers do not have to repeat this
path. The HHL algorithm \cite{harrow2009quantum} essentially performs a complete eigen-decomposition
of a matrix to express the solution, at least in state space, as a linear combination of
eigenvectors.
In Quantum Singular Value Transformation (QSVT)
\cite{martyn2021grand, gilyen2019quantum, dong2021efficient} 
the matrix is directly inverted using an operator that approximates $1/x$ and is modified when
$|x| < 1/\kappa$ to remove the singularity at $x=0$.
Coupled with an implicit CFD solver, direct quantum matrix solvers such as HHL and QSVT are the
most likely to achieve quantum advantage, due to larger memory spaces, faster convergence rates, 
and the fact that only the CFD discretisation and matrix assembly remain on the classical computer. 
In time, there may be quantum equivalents of these too.

This work applies an emulated HHL circuit to an implicit CFD solver, extending the
author's previous HHL assessment for the SIMPLE pressure correction equation.
The test matrices are small enough that they can be compared to a classical eigen-decomposition
obtained using the Gnu Scientific Library (GSL) \cite{gough2009gnu}. 
This allows the effect of the number of qubits in the HHL eigenvalue register to be
understood in terms of the eigenvalues and corresponding eigenvectors that are resolved.
The unresolved components are, in turn, related to the iterative performance of the CFD solver
using established classical eigenvalue theory.
The work is presented as follows. \Cref{sec-implicit-CFD} gives a brief overview of the
implicit CFD method used in this study. \Cref{sec-testcase} describes the CFD test case.
\Cref{sec-solver} briefly describes the emulated hybrid solver, 
for full details see \cite{lapworth2022hybrid}.
\Cref{sec-results} presents the main results and \Cref{sec-conclude} draws conclusions.

%
\section{Implicit CFD}
\label{sec-implicit-CFD}

As in \cite{lapworth2022hybrid}, the steady 2-dimensional incompressible
Navier-Stokes equations are considered, which can be written in component
form as:

\begin{equation}
  \begin{array}{rcl}
    \frac{\partial \rho u}{\partial x} + \frac{\partial \rho v}{\partial y} &=& 0 \\
    \frac{\partial \rho uu}{\partial x} + \frac{\partial \rho vu}{\partial y} &=&  
          \frac{\partial}{\partial x} \left( \mu \frac{\partial u}{\partial x}\right) +
          \frac{\partial}{\partial y} \left( \mu \frac{\partial u}{\partial y}\right) -
          \frac{\partial p}{\partial x} \\
    \frac{\partial \rho uv}{\partial x} + \frac{\partial \rho vv}{\partial y} &=&  
          \frac{\partial}{\partial x} \left( \mu \frac{\partial v}{\partial x}\right) +
          \frac{\partial}{\partial y} \left( \mu \frac{\partial v}{\partial y}\right) -
          \frac{\partial p}{\partial y} \\
  \end{array}
  \label{eqn-navier-2d}
\end{equation}

Using the SIMPLE (Semi-Implicit Method for Pressure Linked Equations) algorithm
\cite{patankar1972calculation}, \cite{ghia1977study} leads to a set of 
segregated matrix equations of the form \cite{versteeg2007introduction}:

\begin{equation}
  \begin{array}{rcl}
   (A^u + B^u) \ket{u^*} &=& D^x \ket{p^*} + \ket{u_b} \\
   (A^v + B^v) \ket{v^*} &=& D^y \ket{p^*} + \ket{v_b} \\
    A^p \ket{p^{\prime}} &=& M^x \ket{u^*} + M^y \ket{v^*} \\
  \end{array}
  \label{eqn-simple-01}
\end{equation}

Where
\begin{itemize}
    \item $A^u$ and $A^v$ are matrices containing the discrete convection and
    diffusion operators for the $u$ and $v$ momentum equations.
    \item $B^u$ and $B^v$ are matrices for the discrete $u$ and $v$ boundary condition operators.
    \item $D^x$ and $D^y$ are the discrete pressure gradient operators.
    \item $A^p$ is the Poisson-like operator for the pressure correction matrix.
    \item $M^x$ and $M^y$ are the components of the discrete continuity operator.
    \item $\ket{u_b}$ and $\ket{v_b}$ are the specified boundary values and have zero amplitudes at the interior nodes.
\end{itemize}

The pressure correction equation has an elliptic character and is more
computationally intensive to solve than the momentum equations which are
hyperbolic. Solving the pressure correction equation on a quantum computer
may address the dominant time factor of each iteration, but the inefficiencies
of a segregated solver remain.

The implicit approach combines the discrete Navier-Stokes equations
into a single matrix equation: 

\begin{equation}
  \begin{pmatrix}
    A^u + B^u &  0        & -D^x \\
    0         & A^v + B^v & -D^y \\
    M^x       & M^y       &  0   \\
  \end{pmatrix}
  \begin{pmatrix}
    \ket{u^{\prime}} \\
    \ket{v^{\prime}} \\
    \ket{p^{\prime}}
  \end{pmatrix}
  =
  \begin{pmatrix}
    R^u \\
    R^v \\
    R^p
  \end{pmatrix}
  \label{eqn-implicit-01}
\end{equation}

Where $R$ terms are the residuals of each of the equations. 
These give the error by which the equations are satisfied and tend to zero as
the calculation converges. The system is solved for corrections, 
$\ket{u^{\prime}}$ etc., which also tend to zero.
The terms $A$ and $R$ depend non-linearly on the flow field and, hence,
\cref{eqn-implicit-01} represents a linearisation about the current
non-linear solution. These schemes are sometimes called {\it coupled}
as the Navier-Stokes equations are solved as a single system.
\Cref{eqn-implicit-01} uses the same fixed point Picard linearisation as used
by \cite{lapworth2022hybrid}, but other linearisations are possible.

Implicit schemes converge more rapidly than segregated
schemes, \cite{mazhar2016novel} report a speed-up of a factor of 20 over SIMPLER, a derivative of
SIMPLE, on a similar test case and mesh size to the one used here.
However, implicit schemes require larger matrices and are less iteratively stable, particularly
in the early iterations, than segregated schemes.
As in \cite{puyero1997efficient} and \cite{blais2020lethe} this instability is often addressed using the
GMRES \cite{saad1986gmres} algorithm which further adds to the memory overhead.
These overheads limit the use of implicit schemes for large scale classical calculations.
If quantum algorithms can be shown to avoid these limitations, implicit solvers
are a strong candidate for quantum advantage.

\begin{figure}[h]
  \begin{center}
    \includegraphics[width=0.5\textwidth]{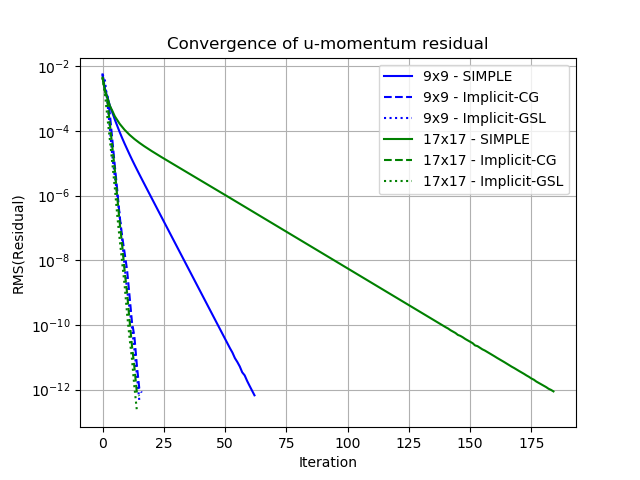}
  \end{center}
  \caption{Comparison of SIMPLE and Implicit convergence histories for
           9x9 and 17x17 meshes ($Re=100$).}
  \label{fig:simple-vs-implicit}
\end{figure}

\Cref{fig:simple-vs-implicit} demonstrates the benefits of an implicit solver
over a segregated one for the test case discussed in \Cref{sec-testcase}.
Convergence histories are shown for two meshes, 9x9 and 17x17, for the
case with Reynolds number $Re=100$.
The solid lines are from the SIMPLE solver and show the number of iterations
being highly dependent on the mesh size. The pressure correction solver 
uses relaxation factors of 0.7 and 0.5 for the velocity and pressure
respectively - these are close to the optimal values. The linearised 
pressure correction and momentum equations are solved to an error 
of less than $10^{-12}$ using a conjugate gradient (CG) solver.
The dashed lines show the convergence of the implicit solver with both
relaxation factors set to 1.0 and the linearised coupled equations also
solved to $10^{-12}$ using a CG solver.
The dotted lines show the classical eigen-solution \cite{golub2013matrix}
using the GNU scientific library (GSL) \cite{gough2009gnu} to 
solve for the eigen-decomposition of the coupled matrix.
In both cases, the implicit solver convergences far more rapidly and
the convergence is independent of the mesh size. This is also the case
for the 33x33 mesh (not shown) where SIMPLE requires 786 iterations and
the implicit solver using CG requires 15 iterations.

Using GSL to solve the coupled matrix illustrates the time complexity of
classical eigen-decomposition. For the 17x17 mesh, the full implicit
solution using CG took 0.06s. whereas GSL took over 24min.
Both were run on the same 
Intel\textsuperscript{\tiny\textregistered}
Core\textsuperscript{\tiny\textcopyright}
i9 12900K 3.2GHz processor.

%
\section{Test case}
\label{sec-testcase}

The lid driven cavity is one of the canonical test cases for CFD with the first published
applications dating from the 1970s \cite{smith1975comparative, ghia1977study}
and with reference solutions provided by \cite{ghia1982high}.
This case continues to be used as a basic validation case for viscous
incompressible flow solvers on coarse meshes \cite{redal2019dynamfluid}.

\begin{figure}[h]
  \begin{center}
    \includegraphics[width=0.60\textwidth]{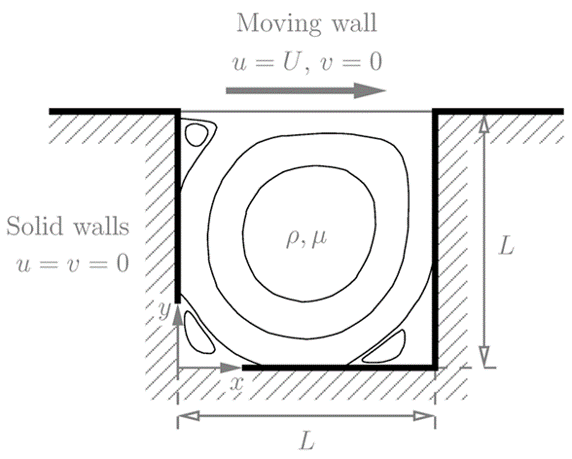}
  \end{center}
  \caption{\centering Overview of lid driven cavity test case, from \cite{redal2019dynamfluid}.}
  \label{fig-cavity-flow}
\end{figure}

Whilst, this work considers only low Reynolds number laminar flow, at higher Reynolds 
numbers this test case exhibits more complicated phenomena such as recirculations,
turbulent flow structures and laminar to turbulent transition.
\Cref{fig-cavity-flow} from \cite{redal2019dynamfluid} gives an indication of some of the
2 dimensional flow structures.
In 3D, the flow demonstrates
complex unsteady turbulence phenomena such as inhomogeneous turbulence and small 
scale helical structures \cite{bouffanais2007large, shetty2010high}.
\cite{shetty2010high} used a staggered mesh and a derivative of the SIMPLE scheme on a
$64 \times 64 \times 64$ mesh.
\cite{courbebaisse2011time} used the equivalent of a $129 \times 129 \times 129$ mesh to 
perform direct numerical simulations of the turbulent flow in a 3D cavity.
Calculations of this type are attractive for quantum computing as they explicitly resolve 
all the turbulence phenomena and do not require complicated and highly non-linear
physical models of turbulence.

\begin{table}[h]
  \centering
  \begin{tabular}{c c c c c c c}
    \toprule
    CFD mesh &  matrix rank & \#non-zeros & $|\lambda_{min}|$    & $|\lambda_{max}|$ & $\kappa$ & HHL \\
    \midrule
     5x5     &   76         & 265         & $3.53\times10^{-2}$ & 0.728          & 20.7     & 14 (7 + 6 + 1)   \\
     9x9     &   244        & 1,105       & $5.72\times10^{-3}$ & 0.415          & 72.6     & 18 (9 + 8 + 1)  \\
     17x17   &   868        & 4,513       & $5.22\times10^{-4}$ & 0.252          & 482      & 23 (11 + 11 + 1)  \\
     33x33   &   3,268      & 18,241      & $3.59\times10^{-5}$ & 0.137          & 3,806    & 27 (13 + 13 + 1)  \\
    \bottomrule \\
  \end{tabular}
  \caption{\centering Dimensions, eigenvalue range and condition number for implicit $Re=100$ matrix.
           The HHL column gives an estimate of the number of logical qubits needed by HHL 
           (state + eigenvalue + ancilla qubits).}
  \label{tab-testcases}
\end{table}

\begin{figure}[h]
  \begin{center}
    \includegraphics[width=0.50\textwidth]{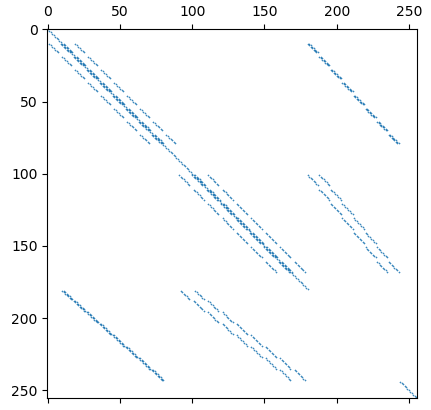}
  \end{center}
  \caption{Sparsity pattern for 9x9 coupled system extended to have rank 256.}
  \label{fig:9x9_sparsity}
\end{figure}

\Cref{tab-testcases} gives the characteristics of the implicit matrices for
a range of mesh sizes. There are two factors of note: (a) the matrix ranks are
not a power of 2; and, (b) the eigenvalues are calculated for the Hermitian matrix
used by HHL and hence are real-valued. The eigenvalues occur in pairs of the
same value with opposite signs.
In order for the system to be amenable to HHL, the matrices are extended by a
diagonal block in the bottom left corner of the matrix to give a rank equal to
the next largest power of 2. This is illustrated in \Cref{fig:9x9_sparsity} which
shows the sparsity pattern for the 9x9 mesh with a diagonal extension to
give a rank of 256.
The added diagonal entries are all set according to the CFD mesh spacing to give
terms with the same order as the other matrix elements.

For the incompressible, laminar flows, as studied here, the only non-linear term 
is the fluid convection term. For the cases with a Reynolds number of 100,
the convection term is only dominant very near to the moving lid. Hence, there
are only minor variations in the eigenvalues of the linearised matrices.
The rapid convergence of the implicit solver also ensures that after the first
2-3 iterations, variations in the eigenvalues are in the second or third
significant figures. \Cref{subsec-results-re} considers higher Reynolds 
numbers where the effect of a greater initial variation in the eigenvalues is
discussed.

Also of interest in \Cref{tab-testcases} is that the condition numbers are significantly
lower than for the pressure correction matrices reported by \cite{lapworth2022hybrid}.
For example, the condition number for the pressure correction matrix on the 17x17 mesh
is 3,500 compared to 482 for the implicit matrix. This means that whilst the rank of the
implicit matrix is 4 times larger, the condition number is 8 times lower and the HHL 
emulations reduce by 1 qubit.

Unless otherwise stated, all calculations are for a Reynolds number of 100.

%
\section{Hybrid CFD solver}
\label{sec-solver}

Since the implicit matrices, $A$, listed in \Cref{tab-testcases} are non-symmetric, they must be
{\it symmetrised} to create a Hermitian operator for the  Quantum Linear Equation Solver (QLES):

\begin{equation}
  H =
  \begin{pmatrix}
    0           & A \\
    A^{\dagger} & 0
  \end{pmatrix}
  \label{eqn-A2H}
\end{equation}

And the linear system to be solved becomes:
\begin{equation}
H 
  \begin{pmatrix}
    0 \\
    x
  \end{pmatrix}
  =
  \begin{pmatrix}
    b \\
    0
  \end{pmatrix}
  \label{eqn-Hx=b}
\end{equation}

Matrix preparation for the HHL circuit requires the matrix $H$ to be decomposed into
a linear combination of $M$ unitaries (LCU):

\begin{equation}
  H = \sum_{i=1}^{M} \alpha_i U_i
  \label{eqn-lcu01}
\end{equation}

\begin{figure}[h]
  \begin{center}
    \includegraphics[width=0.80\textwidth]{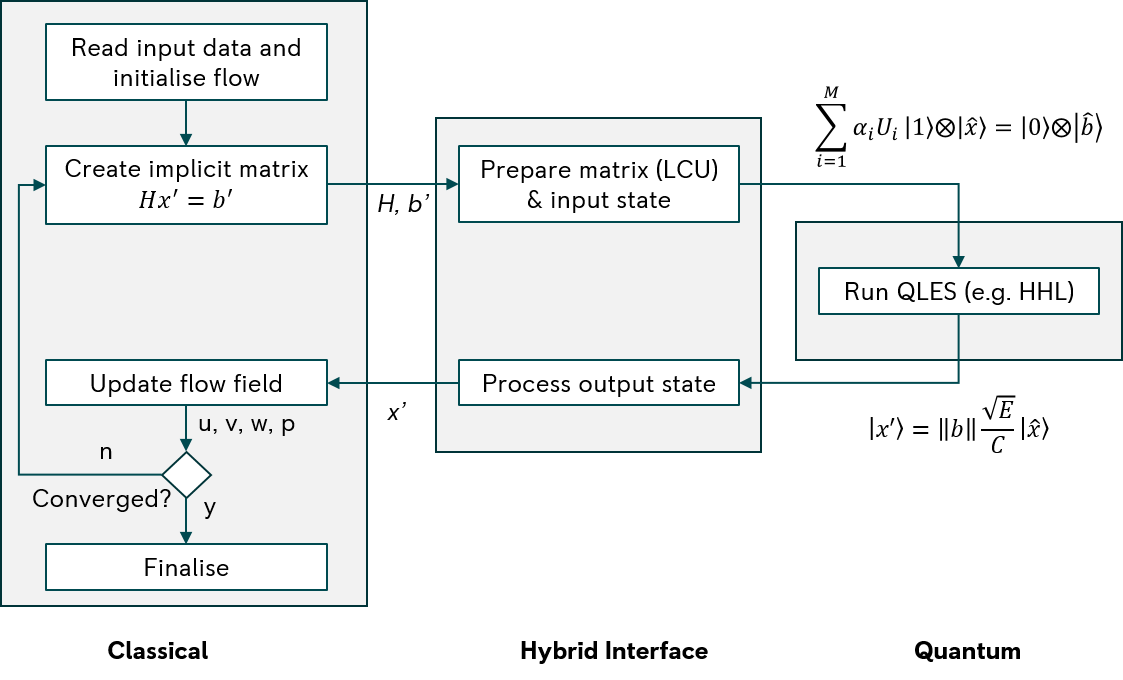}
  \end{center}
  \caption{Hybrid implicit CFD solver. Hats denote normalised variables.}
  \label{fig-hybrid-cfd}
\end{figure}

\begin{figure}[h]
  \begin{center}
    \includegraphics[width=0.80\textwidth]{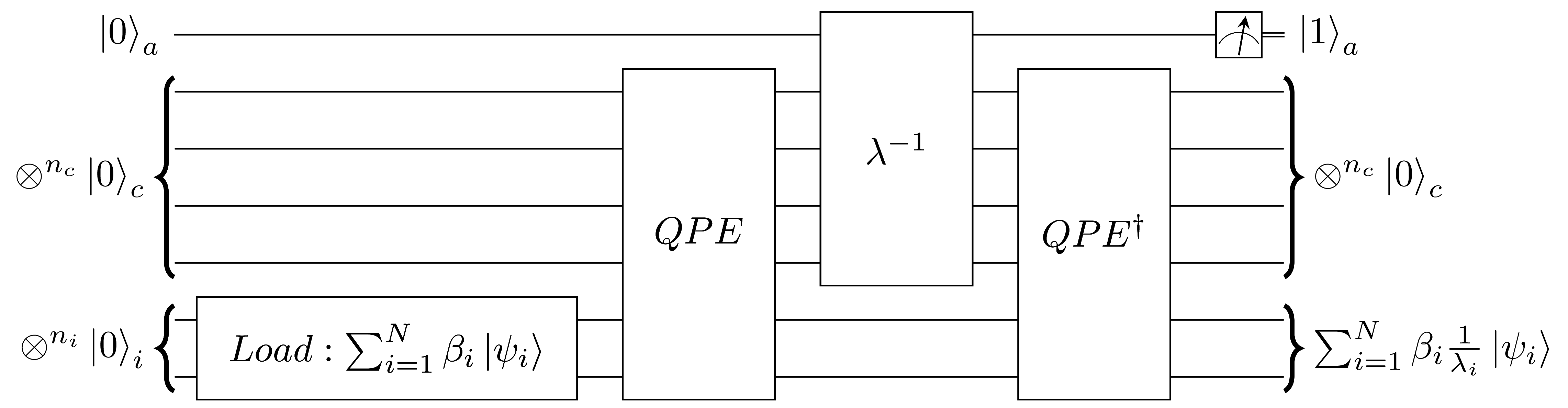}
  \end{center}
  \caption{\centering Outline of HHL circuit consisting of 3 registers: input register (i),
           clock register for eigenvalues (c), ancilla register (a).}
  \label{fig-hhl_01}
\end{figure}

The LCU decomposition takes place on the classical computer and, as shown by \cite{lapworth2022hybrid},
can become the dominant computational cost of a hybrid solver. 
\Cref{fig-hybrid-cfd} gives an overview of the hybrid implicit CFD solver and shows the benefit of
placing all the matrix inversions on a quantum device. 
In this work, the QLES is emulated on a classical computer using the circuit shown in \Cref{fig-hhl_01}.
The emulated HHL solver is as in \cite{lapworth2022hybrid} and is not repeated here.

It will be instructive later to refer to the linear algebra that underpins the HHL algorithm.
Since H is Hermitian, all the eigenvalues are real and the eigenvectors
form an orthonormal basis for the space of $N$-dimensional vectors.
The right hand side vector can, therefore, be written as:

\begin{equation}
  \ket{b} = \sum_{i=1}^{N} \beta_i \ket{\psi_i}
  \label{eqn-b1}
\end{equation}

The coefficients $\beta_i$ are obtained from:
\begin{equation}
  \beta_i = \bra{\psi_i}\ket{b}
  \label{eqn-beta1}
\end{equation}

The eigenvectors of $H$ are also the eigenvectors of $H^{-1}$
and the eigenvalues of $H^{-1}$ are the inverse of the eigenvalues of $H$.
Hence:
\begin{equation}
  H^{-1} = \sum_{i=1}^{N} \frac{1}{\lambda_i} \ket{\psi_i}\bra{\psi_i}
  \label{eqn-Hinv}
\end{equation}

The solution vector $\ket{x}$ can be written as:
\begin{equation}
  \begin{split}
  \ket{x} & =  H^{-1} \ket{b} \\
          & =  \sum_{i=1}^{N} \frac{1}{\lambda_i} \ket{\psi_i}\bra{\psi_i} 
                \sum_{j=1}^{N} \beta_j \ket{\psi_j} \\
          & =  \sum_{i=1}^{N} \beta_i \frac{1}{\lambda_i} \ket{\psi_i}
  \end{split}
  \label{pdes_eq_1}
\end{equation}

If all the eigenvalues and vectors of $H$ can be found, then solving for $\ket{x}$ is
straightforward. This has always been the case, but classical algorithms for
computing all the eigen-pairs have computational complexity $\mathcal{O}(n^3)$.
By contrast, the HHL algorithm has complexity $\mathcal{O}(log(n)s^2\kappa^2/\epsilon)$
where $s$ is the maximum number of non zero entries per row or
column; $\kappa$ is the condition number of the matrix; and
$\epsilon$ is the precision.
In this study, the matrices are small enough to classically compute their
eigen-spectra.

\begin{figure}[h]
  \centering
  \begin{subfigure}[b]{0.33\textwidth}
      \centering
      \includegraphics[height=6cm]{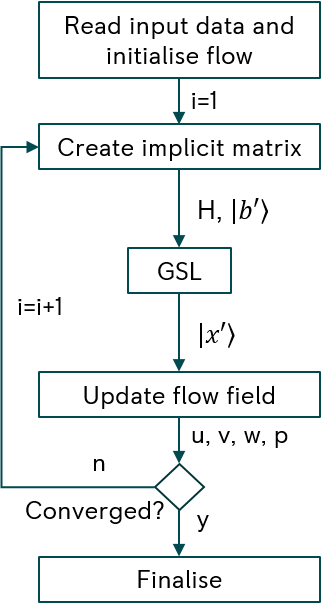}
      \caption{Classical linear solver.}
      \label{fig:GSL_implicit}
  \end{subfigure}
  \hfill
  \begin{subfigure}[b]{0.66\textwidth}
      \centering
      \includegraphics[height=6cm]{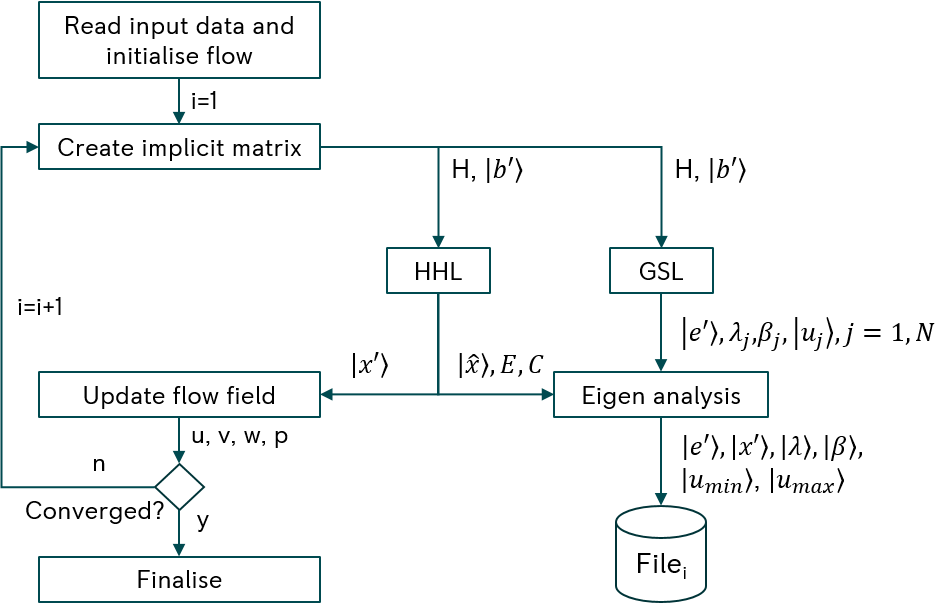}
      \caption{(Verification of HHL linear solutions.}
      \label{fig:GSL_HHL}
  \end{subfigure}
  \caption{\centering Modes of usage for GSL: (a) linear solver for reference fully classical
            solutions, (b) Verifier for each HHL solution using the same input state 
            and matrices as HHL. }
  \label{fig-GSL}
\end{figure}

There are two modes of usage for GSL as shown in \Cref{fig-GSL}.
In the first mode, GSL is used as the linear solver for a completely classical solution.
In the second mode, GSL is used to check the fidelity of the HHL solution for each outer
iteration. GSL takes the same input vector and matrix as HHL and computes the classical
solution for each linear system. The outer non-linear update uses the HHL solution and 
progresses exactly as if the GSL verifier were absent.
For each iteration the following are written to an iteration stamped file: 
the GSL and HHL solutions; all the values of $\lambda$ and $\beta$;
and, the eigenvectors corresponding to the minimum and maximum eigenvalues.

\subsection{Interpreting the eigenvalues of a CFD matrix}
\label{sec-eval-interp}
Prior to presenting the results, it is useful to consider
the physical interpretation of the eigen-system of a CFD matrix.
\Cref{eqn-implicit-01} effectively solves for corrections that
eliminate errors in the satisfaction of the non-linear equations
\Cref{eqn-navier-2d}. The eigenvectors correspond to error waves
propagating in the flow.

The highest eigenvalues correspond to the most energetic errors 
and have eigenvectors with high frequency oscillations between
neighbouring points in the CFD mesh. See the green line in
\Cref{fig:17x17_evecs}. Since the CFD discretisation directly
connects neighbouring points, the errors lead to high residuals
and are rapidly damped by the solver. 
In some discretisation schemes, there is a
velocity-pressure decoupling that leads to a linearised matrix which
has a null space \cite{lien2000pressure}. The null space allows the
so-called pressure checker-board mode to go undamped.
The staggered meshes used in this study have been chosen because
they do not have a null space.

The lowest eigenvalues correspond to eigenvectors with low frequency
oscillations that often have a wavelength equal to the dimension
of the solution domain. These are associated with errors in one
region that affect distant regions via convection or diffusion.
See the blue line in \Cref{fig:17x17_evecs}.
The rationale of the multi-grid method \cite{wesseling1982robust}
is that low frequency errors on
a fine mesh have progressively higher frequencies on the sequence of
coarser meshes. This is why in multi-grid terminology the solver is
often referred to as a {\it smoother}.

Whilst variational solvers and linear equation solvers both result 
in a low energy state in which the $L_2$ norm of the residual errors is low,
the following results show the necessity of knowing and resolving the
full eigenvalue range for HHL.

%
\section{Results}
\label{sec-results}
All results were run on a desktop PC with an 
Intel\textsuperscript{\tiny\textregistered}
Core\textsuperscript{\tiny\textcopyright} i9 12900K 3.2GHz
Alder Lake 16 core processor and 64GB of 3,200MHz DDR4 RAM. 
The calculations were parallelised using OpenMP directives \cite{chandra2001parallel}.
All calculations were performed using double precision arithmetic including the
CFD solver. Whilst the test case does not require this level of precision, it
ensures that trends in the HHL solver are not masked by the precision of
the input state or the matrix.

The main concern of the following analysis is the impact that some of the 
parameters in the linear matrix solver, such as the number of qubits used, have on 
the iterative convergence of the outer non-linear iterations. This is
intimately linked with the accuracy of each inner linear solution but,
as will be shown, there are iterative instabilities that are not apparent
in an individual inner solution. The hybrid solutions are compared with 
{\it exact} solutions obtained using GSL \cite{gough2009gnu} which 
allows a full eigen-spectrum analysis to be performed.
The eigen-spectrum is obtained for the same symmetric matrix, $H$, as used by HHL,
using symmetric bi-diagonalisation and the QR reduction method
(see \S8.3 of \cite{golub2013matrix}). 
The computed eigenvalues are accurate to an absolute accuracy of $\epsilon ||A||_2$, 
where $\epsilon$ is the machine precision.

%
\subsection{Matrix decomposition}
\label{subsec-mat-decomp}

\Cref{tab-lcu-decomp} gives the number of unitaries in the LCU decomposition
for each CFD mesh. The Pauli string approach used by \cite{lapworth2022hybrid} produces a rapidly
increasing number of unitaries to the point that the decomposition was not attempted for
the 17x17 and 33x33 meshes The 13x13 mesh is included to show the rate at which the number
of Pauli strings increases. 

\begin{table}[h]
  \centering
  \begin{tabular}{c c c c c}
    \toprule
    \multicolumn{1}{c}{CFD mesh} & \multicolumn{2}{c}{\#LCU unitaries} \\
                & Pauli strings & Simple 1-sparse & $\kappa$   \\
    \midrule
     5x5        &   4,132      & 634    & 20.7  \\
     9x9        &   16,104     & 2,234  & 72.6  \\
     13x13      &   48,956     & 5,050  & 205   \\
     17x17      &   -          & 9,338  & 482   \\
     33x33      &   -          & 38,138 & 3,806 \\
    \bottomrule \\
  \end{tabular}
  \caption{Number of unitaries in the LCU decomposition of the implicit CFD matrices.}
  \label{tab-lcu-decomp}
\end{table}

For this work, an alternative simple 1-sparse decomposition is used where 
each pair of symmetric
entries in $H$ corresponds to a pair of matrices in the LCU. 
The decomposition begins with 2 matrices $I_{+} = I$ and $I_{-} = -I$
where $I$ is the rank $N$ unit matrix. 
If $h_{ij} = h_{ji}$ are the symmetric entries in $H$, then,
in both $I_{+}$  and $I_{-}$, the entries $ij$ and $ji$ are set to $1$ and the
entries $ii$ and $jj$ are set the $0$.
The coefficient for both matrices is $h_{ij}/2$. 
As \Cref{tab-lcu-decomp} shows the number of unitaries is equal to the number of
non-zero entries in $H$. 
Calculating this decomposition is extremely efficient and enables a broader range of
emulations to be studied. However, the circuit implementation on a physical device
has not been considered and is expected to be significant.

\begin{figure}[h]
  \begin{center}
    \includegraphics[width=0.50\textwidth]{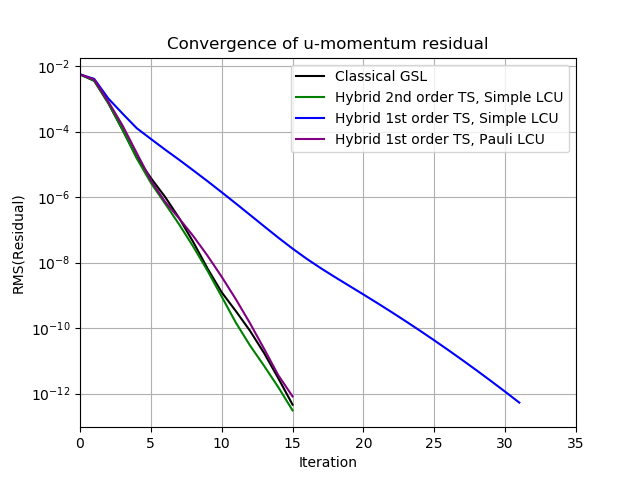}
  \end{center}
  \caption{Effect of Trotter approximation on the convergence of the hybrid implicit solver on 
           the 9x9 CFD mesh.}
  \label{fig:9x9_lcu_comp}
\end{figure}

\Cref{fig:9x9_lcu_comp} compares how the two LCU options affect the
convergence rate on the 9x9 CFD mesh including the order of the
Trotter-Suzuki (T-S) approximation \cite{trotter1959product, suzuki1993improved}
for the Hamiltonian evolution operator.
Both the Pauli string LCU with first order T-S and the simple pairwise LCU with
second order T-S match the exact convergence rate very well. 
Whereas the simple pairwise LCU with first order T-S takes twice as many iterations
to converge, despite having a factor of 8 fewer unitaries in the LCU.

The error orders for the Trotter-Suzuki approximations are given in
\Crefrange{eqn-trotter-01}{eqn-trotter-02} where $\bar{\alpha}$ is an
average LCU coefficient.

\begin{equation}
  e^{-i \sum_{j=1}^{M} \alpha_j U_j t} = \left( \prod\limits_{j=1}^{M} e^{-i \alpha_j U_jt/r} \right)^r + \mathcal O(\bar{\alpha}^2t^2/r)
  \label{eqn-trotter-01}
\end{equation}

\begin{equation}
  e^{-i \sum_{j=1}^{M} \alpha_j U_j t} = \left( \prod\limits_{j=1}^{M} e^{-i \alpha_j U_jt/2r}
                                                  \prod\limits_{j=M}^{1} e^{-i \alpha_j U_jt/2r} \right)^r + \mathcal O(\bar{\alpha}^3t^3/r^2)
  \label{eqn-trotter-02}
\end{equation}

For the Pauli string LCU, the root mean square of the LCU coefficients
is $1.5 \times 10^{-3}$, whereas, for the simple pairwise LCU it is
$4.6 \times 10^{-2}$.
This leads to an almost 3 orders of magnitude difference in leading
error term for the first order scheme which far outweighs the factor
8 increase in the number of unitaries.

An alternative to Trotter-Suzuki is qubitisation
\cite{Low2019hamiltonian, childs2012hamiltonian, kothari2014efficient, berry2015simulating, berry2018improved, babbush2018encoding} 
which provides an exact implementation of a unitary decomposition
at the expense of an additional ancilla register for 
{\it preparing} the coefficients of the unitary decomposition.
For the 9x9 mesh, the prepare register for the Pauli LCU would require 14 qubits
and for the simple pairwise LCU would require 12 qubits. These are both
more than the 8 qubits required by the HHL eigenvalue register (\Cref{tab-9x9-prec}).

Unless otherwise stated, the following results used the second order 
Trotter-Suzuki formula with the simple pairwise LCU.

%
\subsection{Precision}
\label{subsec-precision}

The analysis begins with the 9x9 CFD meshes as both the hybrid emulator and the
{\it exact} GSL calculation can run the full implicit calculations in matters of minutes.
\Cref{tab-9x9-prec} compares the exact eigenvalues with the range of
values that can be resolved with different numbers of qubits in the
eigenvalue register. A precision $(s,m,n)$ indicates $s$ sign qubits, $m$ integer qubits and 
$n$ fraction qubits.
Where $m$ is negative, this indicates a right shift of the binary decimal 
point by $m$ spaces in the output bit patterns.
The exact values are computed using GSL and taken from classical solutions after 10
non-linear iterations.

\begin{table}[h]
  \centering
  \begin{tabular}{c c c c c c c}
    \toprule
    Precision &   $\lambda_{min}$     &  $\lambda_{max}$  & Total HHL qubits\\
    \midrule
     Exact (GSL)  &   $\pm 0.00572$  &  $\pm 0.415$    & -  \\
     (1,-1,6)     &   $\pm 0.01563$  &  $\pm 0.484$    & 16 \\
     (1,-1,7)     &   $\pm 0.00781$  &  $\pm 0.492$    & 17 \\
     (1,-1,8)     &   $\pm 0.00391$  &  $\pm 0.496$    & 18 \\
    \bottomrule \\
  \end{tabular}
  \caption{Exact eigenvalues and resolution for a range of precisions
           for the 9x9 implicit matrix.}
  \label{tab-9x9-prec}
\end{table}

\begin{figure}[h]
  \begin{center}
    \includegraphics[width=0.50\textwidth]{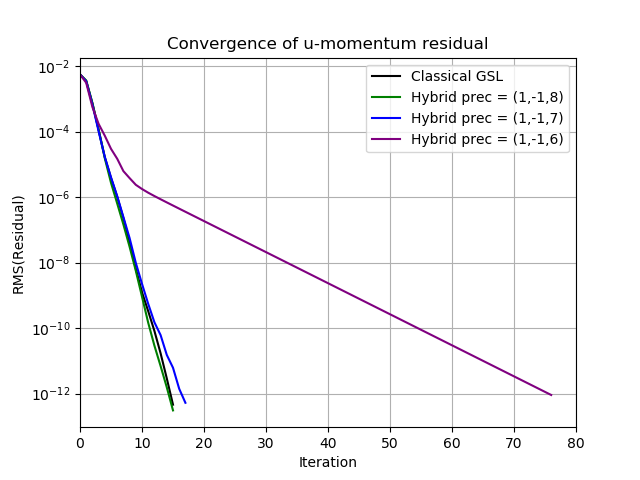}
  \end{center}
  \caption{Effect of fraction precision on the convergence of the hybrid implicit solver on 
           the 9x9 CFD mesh.}
  \label{fig:coupled-9x9-prec1}
\end{figure}

\Cref{fig:coupled-9x9-prec1} compares the convergence of the $u$-momentum
residuals for the 9x9 CFD mesh. Although the implicit solver solves
a single coupled system, it is still instructive to analyse the individual
equations as there can sometimes be iterative feedback mechanisms between the
equations. This is not the case here and the $v$-momentum and continuity 
equations show similar convergence rates. 
The hybrid solution with 8 fraction qubits shows an almost identical
convergence as the classical {\it exact} solution using GSL.
This is as expected as the precision range is 
better than the exact eigenvalue range.
The hybrid solution with 7 fraction qubits is marginally slower by 2
iterations. This is an indication that even a precision that is higher
than the lowest eigenvalue by $37\%$ has an effect on convergence.
The hybrid solution with 6 fraction qubits also converges but takes
almost 5 times as many non-linear iterations. In this case, the precision
is 2.7 times higher than the lowest eigenvalue.

To further investigate the loss of precision, two measures of solution fidelity are used:

\begin{equation}
    f_n = \bra{\hat{x}}\ket{\hat{e}}
    \label{eqn-fnorm}
\end{equation}

\begin{equation}
    f_r = \frac{\bra{x}\ket{e}}{\bra{e}\ket{e}}
    \label{eqn-fraw}
\end{equation}

$f_n$ is the normalised fidelity and compares the state vector at the end of HHL,
$\ket{\hat{x}}$, with the normalised GSL solution, $\ket{\hat{e}}$. 
This is bounded by 0 and 1.
$f_r$ is the {\it raw} fidelity and compares the dimensional solutions normalised
by the exact state. This is not bounded by 0 and 1 but is equal to 1 if the two
solution states match.
$f_n$ and $f_r$ are evaluated for each non-linear iteration using GSL in verification
mode as shown in \Cref{fig:GSL_HHL}.

\begin{figure}[h]
  \begin{center}
    \includegraphics[width=0.50\textwidth]{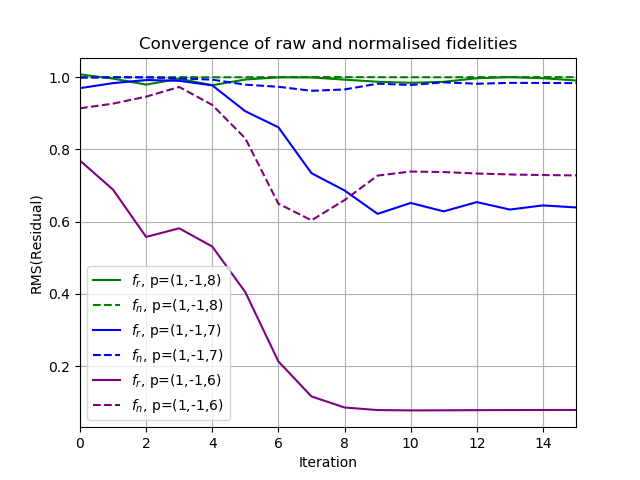}
  \end{center}
  \caption{Comparison of fidelities $f_n$ and $f_r$ for the first 15 hybrid 
           implicit iterations on the 9x9 CFD mesh.}
  \label{fig:9x9_fidelity}
\end{figure}

\Cref{fig:9x9_fidelity} compares these fidelities for the first 15 iterations of
each precision.
Even with (1,-1,8) precision, the raw fidelity shows a small amount of 
variation around 1.0, whilst the normalised fidelity is always above 0.99.
With (1,-1,7) precision, although the normalised fidelity is close to 1.0,
the raw fidelity starts dropping after 4 iterations and stabilises after 10
iterations at about 0.65. Even when the normalised fidelity is close to 1.0 after 
iteration 10, the raw fidelity remains at 0.65.
The (1,-1,6) precision fidelities show the same behaviour as (1,-1,7) but more
exaggerated with the raw fidelity dropping below 0.1 and not recovering.

\begin{figure}[h]
  \centering
  \begin{subfigure}[b]{0.49\textwidth}
      \centering
      \includegraphics[width=\textwidth]{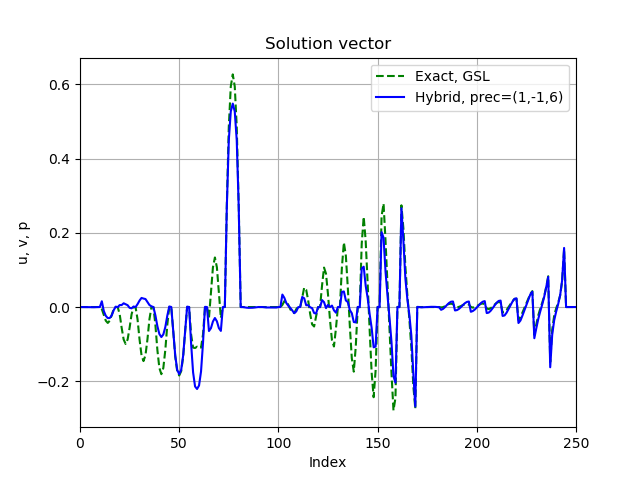}
      \caption{Solutions for the 1st iteration.}
      \label{fig:9x9_1_6_sol_iter01}
  \end{subfigure}
  \hfill
  \begin{subfigure}[b]{0.49\textwidth}
      \centering
      \includegraphics[width=\textwidth]{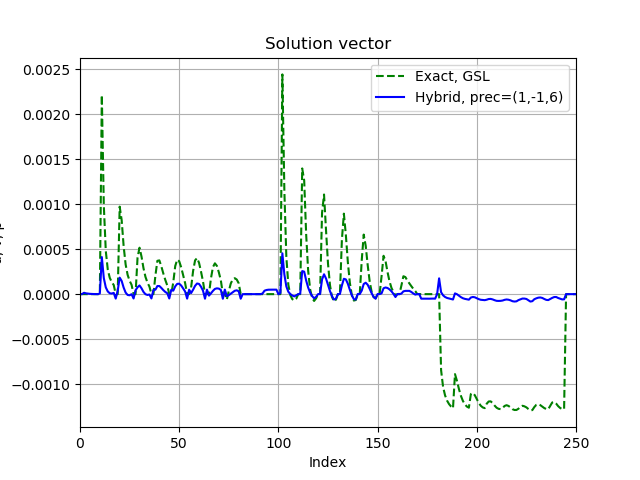}
      \caption{Solutions for the 10th iteration.}
      \label{fig:9x9_1_6_sol_iter10}
  \end{subfigure}
  \caption{\centering Comparison of GSL and (1,-1,6) precision hybrid  
           solutions after 1 and 10 non-linear iterations on the 9x9 CFD mesh.
           Graphs are for dimensionalised solutions which reduce as the non-linear iterations proceed.}
  \label{fig-9x9-1.1.6-comp}
\end{figure}

The significant loss of fidelity in the (1,-1,6) solution between the
first and tenth iterations is illustrated in
\Cref{fig-9x9-1.1.6-comp} which compares the hybrid and GSL verification solutions 
for these iterations.
On the first iteration, there are clear differences between the GSL and
hybrid solutions but the broad scale of the hybrid solution is generally 
correct. The solutions in the range [180, 243], which corresponds to the
pressure field, match very well.
At the tenth iteration, the amplitudes of the hybrid solution are much
smaller than the GSL solution, particularly for the pressure field.

The fact that both solutions in \Cref{fig:9x9_1_6_sol_iter10} 
use the same input vector and matrix, derived from the solution at iteration 9,
suggests that the input vector is dominated by eigen coefficients that the HHL
solution is unable to resolve in the preceding iterations.
Using the GSL eigen decomposition, both solutions can be
compared in terms of the coefficients in their eigen expansions using the
equivalent of \Cref{eqn-beta1}.

\begin{figure}[h]
  \centering
    \begin{subfigure}[b]{0.49\textwidth}
      \centering
      \includegraphics[width=\textwidth]{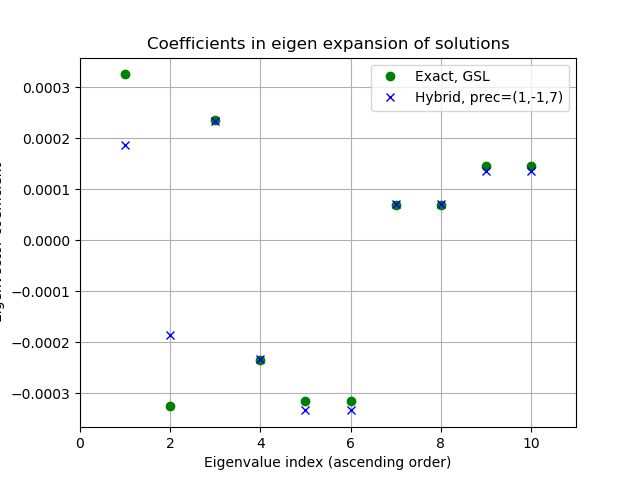}
      \caption{(1,-1,7) precision, $6^{th}$ iteration.}
      \label{fig:9x9_1_7_eigencoeffs}
  \end{subfigure}
  \hfill
  \begin{subfigure}[b]{0.49\textwidth}
      \centering
      \includegraphics[width=\textwidth]{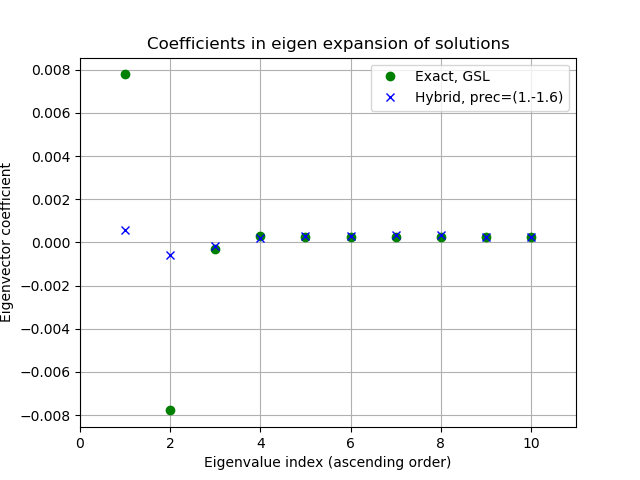}
      \caption{(1,-1,6) precision, $10^{th}$ iteration.}
      \label{fig:9x9_1_6_eigencoeffs}
  \end{subfigure}
  \caption{\centering Coefficients of the eigen-expansion of the GSL and hybrid
            solutions for the first 10 smallest eigenvalues.}
  \label{fig-9x9-eigencoeffs}
\end{figure}

\Cref{fig-9x9-eigencoeffs} compares for eigen coefficients for the first 10
smallest eigenvalues for the (1,-1,6) hybrid solution after 10 outer iterations and
the (1,-1,7) hybrid solution after 6 outer iterations. These solutions were chosen as, 
they have approximately the same level of residual convergence.
The fist four eigenvalues are below the $\pm 0.015625$ precision of 6 fraction qubits and
only the first two are below the $\pm 0.0078125$ precision of 7 fraction qubits.
Note that these comparisons are between two different precision calculations 
at roughly the same level of convergence.

Considering \Cref{fig:9x9_1_7_eigencoeffs}, the coefficients for the (1,-1,7) solution match, 
the exact GSL coefficients very well even for the first eigenvalue pair that 
it has not resolved.
From \Cref{fig:9x9_1_6_eigencoeffs},
the coefficients for the (1,-1,6) solution fail to match the 
exact coefficients for the first eigenvalue pair by an order of magnitude.
Since this is largest coefficient in the eigen-expansion, it explains why the
raw fidelity is so low.

These results are indicative of an iterative feed-forward mechanism at play.
The failure to resolve the lowest eigenvalues creates a constructive reinforcement
on consecutive non-linear iterations whereby 
the residuals corresponding to unresolved eigenvalues are
under corrected. 
Hence, $\beta_0$ and $\beta_1$ in \Cref{eqn-b1} increase 
for successive input vectors, as do
the coefficients of the exact solution $\beta_0/\lambda_0$ and $\beta_1/\lambda_1$.
The disparity increases for about 8 iterations until an equilibrium is reached
and the hybrid solution slowly converges as shown in \Cref{fig:coupled-9x9-prec1}.
The (1,-1,7) solution is subject to the same mechanism but this has a much more 
marginal effect.

Since the eigen-expansions of both the HHL and GSL solutions are available,
the effect of the discrepancy in the first two eigen coefficients can be
used to create a {\it corrected} HHL solution as shown in \Cref{eqn-x-corrected}.

\begin{equation}
  \begin{split}
  \ket{x}_{hhl} & =  \sum_{i=1}^{n} \beta_{x,i} \ket{\psi_i}\bra{\psi_i} \\
  \ket{x}_{cor} & = \sum_{i=1}^{n} \beta_{x,i} \ket{\psi_i}\bra{\psi_i}
                  + \left(\frac{\beta_1}{\lambda_1} -  \beta_{x,1}\right) \ket{\psi_1}
                  + \left(\frac{\beta_2}{\lambda_2} -  \beta_{x,2}\right) \ket{\psi_2}
  \end{split}
  \label{eqn-x-corrected}
\end{equation}

\begin{figure}[h]
  \begin{center}
    \includegraphics[width=0.60\textwidth]{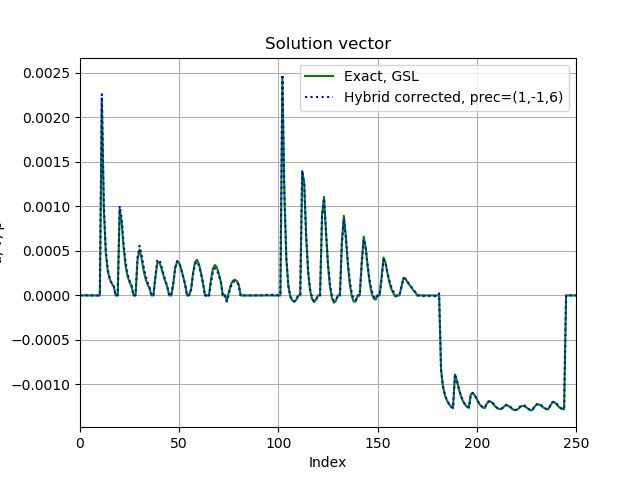}
  \end{center}
  \caption{Comparison of corrected HHL solution from \Cref{fig:9x9_1_6_sol_iter10} 
           using \Cref{eqn-x-corrected} and exact GSL solution.}
  \label{fig:9x9_1_6_sol_iter10_corrected}
\end{figure}

\Cref{fig:9x9_1_6_sol_iter10_corrected} compares the corrected HHL and GSL
solutions after iteration 10 using a precision of (1,-1,6) for HHL.
Whilst there are some tiny discrepancies, the agreement is far better than 
expected.
On a positive note, this illustrates that HHL has accurately represented
all the eigen coefficients that it is able to resolve.
On the negative side, the failure to resolve the lowest eigenvalues has
iteratively driven the calculation to a state where the residual errors, $\ket{b}$,
are dominated by the eigenvectors corresponding to the unresolved eigenvalues.

\subsubsection{Resolving the lowest eigenvalue}
\label{subsubsec-emin}

\begin{figure}[h]
  \begin{center}
    \includegraphics[width=0.60\textwidth]{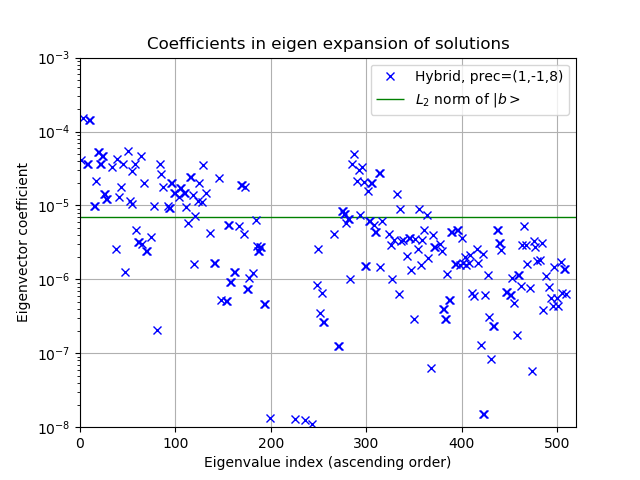}
  \end{center}
  \caption{\centering Coefficients, $\beta$, of the eigen-expansion of the (1,-1,8) solution 
           on the 9x9 CFD mesh after 6 iterations. Green line shows the
           $L_2$ norm of $\ket{b}$.}
  \label{fig:19x9_1_8_eigencoeffs_all}
\end{figure}

The above analysis shows that a good estimation of the lowest eigenvalue
is essential for HHL to be used within an outer non-linear iterative solver.
Even an error of less than a factor of 2 is sufficient to degrade the rate
of non-linear convergence. However, if HHL were able to deliver an exponential
speed-up, a linear degradation in the outer rate of convergence would not remove the
quantum advantage.
There are a number of challenges to obtaining exponential speed-up with HHL.
Not least, the depth of the eigenvalue inversion circuit. 
\Cref{fig:19x9_1_8_eigencoeffs_all} shows all 512 $\beta$ coefficients in the 
eigen-expansion of the output vector on the 9x9 mesh after 6 iterations.
Also shown is a line corresponding the $L_2$ norm of the input vector.
Well over half of the coefficients are within an order of magnitude of
the $L_2$ norm and, possibly, the majority are needed to get an accurate solution.
A logarithmic depth circuit would contain 9 rotations and, hence, would accurately
resolve only 9 of the 512 eigen-coefficients.

Assuming an efficient eigenvalue inversion circuit is available,
the estimation of lowest eigenvalue is an input to the circuit design and,
currently, must be calculated classically.
The power method (see \S7.3.1 of \cite{golub2013matrix}) provides 
a way to estimate the maximum eigenvalue using only matrix-vector multiplication.
The power method can be applied to the inverse of the matrix, $H^{-1}$ to 
find the minimum eigenvalue.
However, inverting $H$ is equivalent to solving the linear system.
Whilst this is potentially a one-off initial calculation, or one that is only 
repeated a small number of times, it is classically intractable for the
size of matrices at which quantum advantage might be expected.

\subsubsection{Resolving the largest eigenvalue}
\label{subsubsec-emax}

The 17x17 coupled matrix has a maximum eigenvalue of 0.2519 which is
very close to a binary precision of $\frac{1}{4}$ and, hence, suitable for
examining its resolution. 
\Cref{tab-17x17-prec} shows the precision ranges for (1,-1,11) and
(1,-2,11). Both fully resolve the lowest eigenvalue and the (1,-2,11)
precision is within $1\%$ of the largest eigenvalue.

\begin{table}[h]
  \centering
  \begin{tabular}{c c c c c c c}
    \toprule
    Precision &   $\lambda_{min}$     &  $\lambda_{max}$  & Total HHL qubits\\
    \midrule
     Exact (GSL)  &   $\pm 0.000522$  &  $\pm 0.2519$    & -  \\
     (1,-1,11)    &   $\pm 0.000488$  &  $\pm 0.4995$    & 23 \\
     (1,-2,11)    &   $\pm 0.000488$  &  $\pm 0.2495$    & 22 \\
    \bottomrule \\
  \end{tabular}
  \caption{Exact eigenvalues and resolution for a range of precisions
           for the 17x17 implicit matrix.}
  \label{tab-17x17-prec}
\end{table}

\begin{figure}[h]
  \centering
  \begin{subfigure}[b]{0.49\textwidth}
      \centering
      \includegraphics[width=\textwidth]{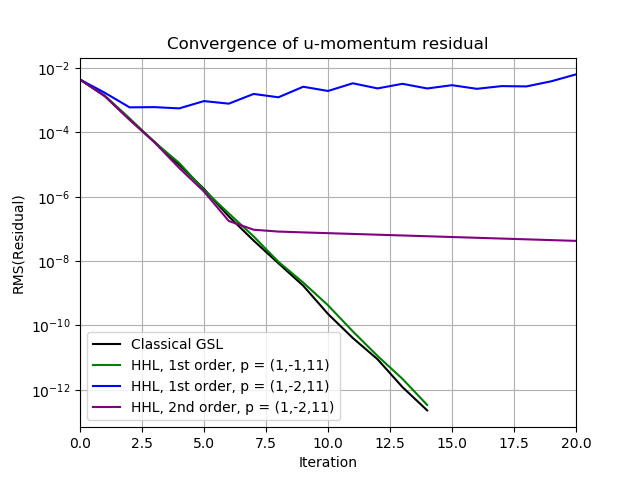}
      \caption{Convergence histories.}
      \label{fig:17x17-conv-prec}
  \end{subfigure}
  \hfill
  \begin{subfigure}[b]{0.49\textwidth}
      \centering
      \includegraphics[width=\textwidth]{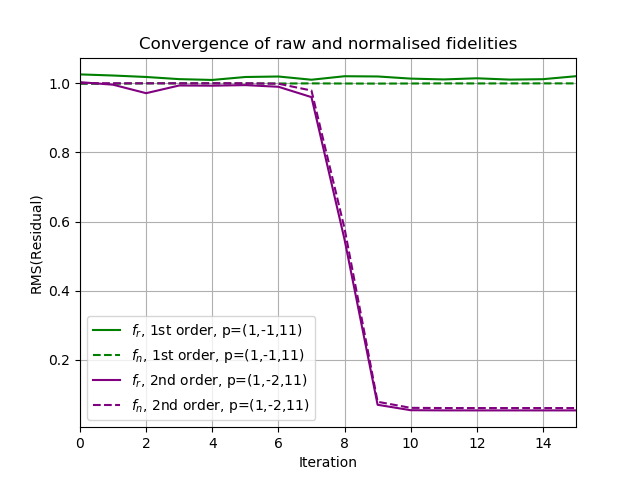}
      \caption{(Raw and normalised fidelities.}
      \label{fig:17x17_fidelity}
  \end{subfigure}
  \caption{\centering Effect of precision and Trotter-Suzuki order for 17x17 CFD mesh.}
  \label{fig-17x17-conv}
\end{figure}

\Cref{fig:17x17-conv-prec} compares the GSL convergence histories with
the hybrid HHL solutions for both precisions and for different Trotterisation orders.
The (1,-1,11) emulation spans the full range of eigenvalues and matches the
exact GSL convergence with the first order Trotter scheme. 
However, changing the integer precision has a secondary effect, that does not
happen with the fraction precision, which is it to change the evolution time in
the quantum phase estimation (QPE) step.
In these emulations, the time-step must be consistent with bit resolution of the
eigenvalue register and is set as:

\begin{equation}
  t = \frac{2\pi}{2^{s+i}}
\end{equation}

where, as before, $s$ is the number of sign qubits and $i$ is the number of fraction 
qubits. Hence, reducing the integer precision by one qubit doubles the QPE
evolution time. 

In both the first and second order simulations the exponent is $r=5$ and
the mean LCU coefficient is approximately $\bar{\alpha}=0.02$.
For the (1,-2,11) precision, the evolution time step is $4\pi$.
From \Crefrange{eqn-trotter-01}{eqn-trotter-02} this gives
a leading error of order $10^{-2}$ for the first order scheme and 
$10^{-4}$ for the 2nd order scheme.

The failure to converge of the first order (1,-2,11) solution in \Cref{fig:17x17-conv-prec}
is due to the Trotter approximation and not the loss of precision.
The second order solution initially converges almost identically to the exact solution,
reducing the residual by almost 5 orders of magnitude, before stagnating from
iteration 6 onward. 
The 2nd order fidelities in \Cref{fig:17x17_fidelity} show a different character from
those for the minimum eigenvalue, 
with both the raw and normalised fidelities moving in concert.

\begin{figure}[h]
  \centering
  \begin{subfigure}[b]{0.49\textwidth}
      \centering
      \includegraphics[width=\textwidth]{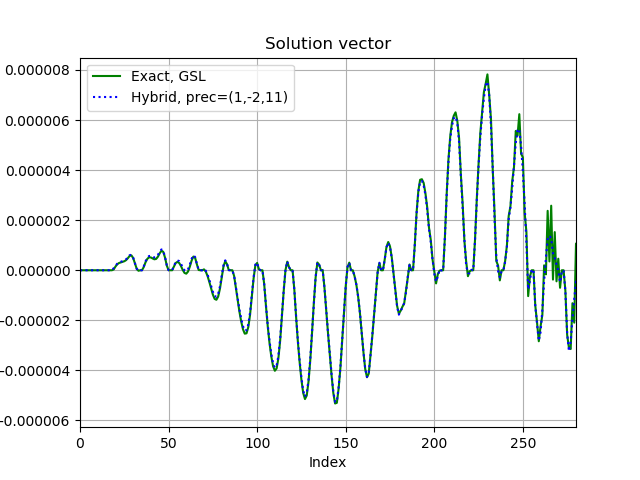}
      \caption{Iteration 8.}
      \label{fig:17x17_-2_11t2_iter08}
  \end{subfigure}
  \hfill
  \begin{subfigure}[b]{0.49\textwidth}
      \centering
      \includegraphics[width=\textwidth]{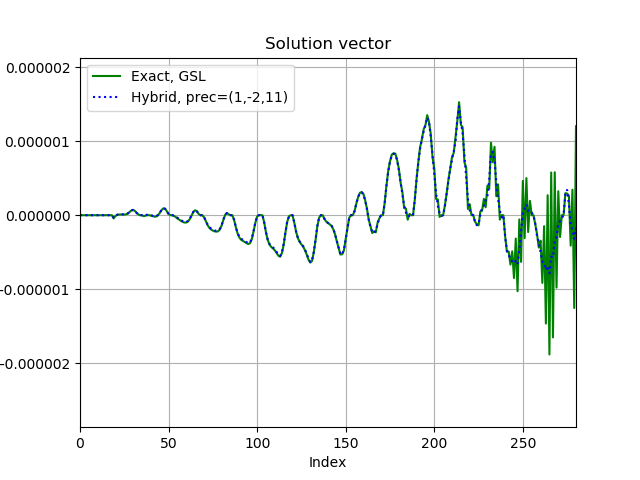}
      \caption{(Iteration 9.}
      \label{fig:17x17_-2_11t2_iter09}
  \end{subfigure}
  \caption{\centering Solutions for the u-momentum equation at iterations 8 and 9
           for 17x17 CFD mesh.}
  \label{fig-17x17-iter8-9}
\end{figure}

\Cref{fig-17x17-iter8-9} shows the solutions for the u-momentum part of the 
state vector after iterations 8 and 9. Small high frequency oscillations near
index 255 at iteration 8, grow rapidly at iteration 9 and by iteration 10 (not shown)
dominate the solution. There are no oscillations present at iteration 6 and only
very minor oscillations at iteration 7.
Note also that the oscillations are in the GSL verification solution and not the hybrid solution.
These have built up in the input vector as before.

\begin{figure}[h]
  \begin{center}
    \includegraphics[width=0.60\textwidth]{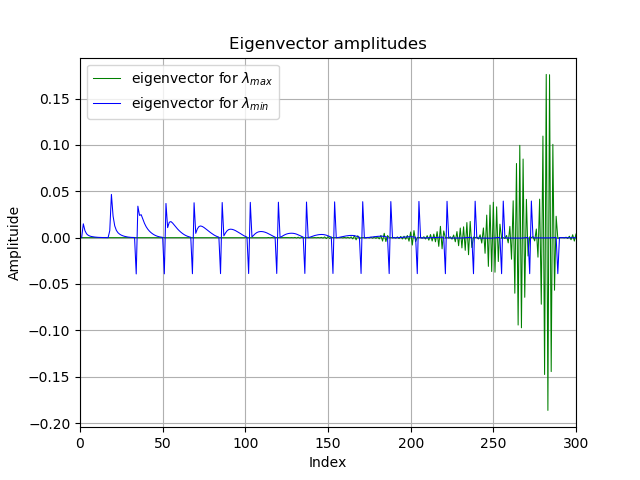}
  \end{center}
  \caption{\centering Amplitudes [0,300] of the eigenvectors for the minimum and maximum
           eigenvalues on the 17x17 CFD mesh.}
  \label{fig:17x17_evecs}
\end{figure}

\Cref{fig:17x17_evecs} compares the first 300 amplitudes of the eigenvectors for
the minimum and maximum eigenvalues. The patterns largely repeat for 
the remaining amplitudes.
As can be seen, the eigenvector for the largest eigenvalue corresponds to a high
frequency oscillation between neighbouring grid points. 
As described in \Cref{sec-eval-interp}, mesh based CFD algorithms
are designed to eliminate, or {\it smooth}, such oscillations very quickly.
By failing to resolve the largest eigenvalue, the highest frequency eigenvector
is not resolved and the oscillations that it would eliminate are allowed to
persist. The character of a sudden switch at between iterations 7 and 9 is due to the
small 1\% difference between the eigenvalue and the precision.
The oscillations begin on the first outer iteration 
at a level of $10^{-12}$ and grow until they reach
$10^{-7}$ at which point they dominate the input and output vectors.
If the precision difference were larger, then a more immediate effect would be expected.

As described in \Cref{sec-eval-interp}, the minimum eigenvalue corresponds
to the eigenvector
for the lowest frequency oscillations which CFD algorithms take many
iterations to eliminate and why acceleration schemes such as multi-grid are
successful. The spikes in the eigenvector amplitudes correspond to 
discontinuities where the mesh index flips from one side wall of the cavity
to the other.

\subsection{Reynolds number}
\label{subsec-results-re}

The dimensions of the CFD mesh are not the only factors that influence
the eigen-spectrum of the implicit matrix.
The other important factor is Reynolds number defined by:

\begin{equation}
    Re = \frac{\rho u L}{\mu}
\end{equation}

Where $L$ is a representative length scale.
For the cavity, $L$ is the cavity width, $u$ is the velocity of the
lid, the density $\rho$ is unity and the viscosity $\mu$ is set to give
the required value of $Re$.

\begin{table}[h]
  \centering
  \begin{tabular}{c c c c c c c}
    \toprule
    Reynolds number &   $\lambda_{min}$     &  $\lambda_{max}$  & $\kappa$ & HHL precision\\
    \midrule
     100     &   $\pm 5.72\times10^{-3}$  &  $\pm 0.415$    & 72.6 & (1,-1,8)  \\
     1,000   &   $\pm 2.11\times10^{-3}$  &  $\pm 0.363$    & 172  & (1,-1,9)  \\
     10,000  &   $\pm 5.10\times10^{-4}$  &  $\pm 0.352$    & 691  & (1,-1,11)  \\
    \bottomrule \\
  \end{tabular}
  \caption{\centering Exact eigenvalues and resolution for a range of precisions
           for the 9x9 implicit matrix after 10 non-linear iterations.}
  \label{tab-9x9-rey}
\end{table}

\Cref{tab-9x9-rey} gives the eigenvalues and qubit counts for Re=100, 1,000 and 10,000
for the 9x9 CFD mesh. A Reynolds number of 10,000 is the point at which the flow is expected to
transition to turbulence. Each precision fully spans the corresponding
eigenvalue range.

\begin{figure}[h]
  \begin{center}
    \includegraphics[width=0.60\textwidth]{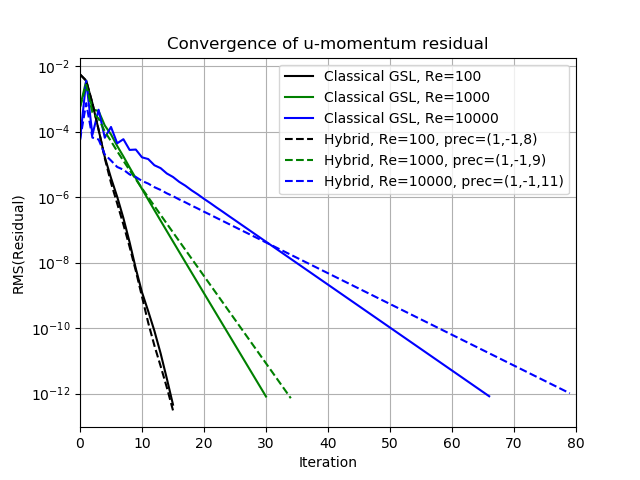}
  \end{center}
  \caption{\centering Effect of Reynolds number on the convergence of the 
           hybrid implicit solver on the 9x9 CFD mesh.}
  \label{fig:9x9-conv-rey-prec}
\end{figure}

\Cref{fig:9x9-conv-rey-prec} compares the GSL and hybrid convergence
histories for the three Reynolds numbers. There is a shared trend for the
number of non-linear iterations to increase with Reynolds number.
This is because the higher the Reynolds number, the higher the
non-linear inertial effects are in the fluid. The solution of the
linearised inner equations is then a less good approximation to the final 
non-linear solution.

There is also a trend that the hybrid solver takes an increasing number
of additional iterations relative to GSL as the Reynolds number increases.
This was traced to the initial iterations for the Re 1,000 and 10,000 having
lower minimum eigenvalues than expected. The numbers in \Cref{tab-9x9-rey}
are taken from the GSL solutions after 10 iterations. 
For the Re=1,000 case this gave a low raw fidelity of 0.67 for the first 
iteration.
For the Re=10,000 case, the raw fidelities ranged from 0.23 to 0.66 for the
first 3 iterations. After that, both cases consistently had raw fidelities 
close to 1.
As shown previously, unresolved solution components in the output state can
transfer to the next input state and slow convergence.
This was confirmed by running the Re=1,000 case with an additional 2 fraction
qubits. The raw fidelities were above 0.98 for all iterations and the
hybrid solver took just 1 more iteration that the GSL solution.

%
\section{Conclusions}
\label{sec-conclude}

Quantum algorithms for direct matrix inversion, such as HHL and QSVT, mean that quantum
CFD algorithms should not need to follow the same path as classical solvers towards highly
sophisticated iterative solvers such as those based on Krylov sub-spaces.
Coupled with implicit discretisation schemes, these provide a promising route to
Quantum Advantage, albeit on Fault Tolerant devices.

This work has demonstrated the importance of assessing HHL within the context of an
outer non-linear iteration rather than simply as linear solver of a individual 
matrix equation. The precision with which HHL resolves both the minimum and maximum 
eigenvalues has been shown to be crucial to how it behaves within a non-linear CFD
solver. The behaviour has been shown to be consistent with how the eigenvalues and
eigenvectors relate to the CFD discretisation.
These findings are relevant to early Fault Tolerant CFD prototypes where every (logical) qubit is likely to count. 

The importance of accurately estimating the eigenvalue range, particularly the
maximum eigenvalue, has been illustrated. This is part of the classical computing
overhead for hybrid algorithms discussed by \cite{lapworth2022hybrid} for the
pressure correction equation. The matrix decomposition overhead using Pauli strings
is more severe for the implicit method with the 9x9 CFD matrix decomposition
containing 50 times more unitaries than the pressure correction matrix.
A simple alternative has been used here, but this is unlikely to lead to
efficient circuits. Although QSVT matrix inversion has not been studied, the
observations on eigenvalue accuracy (i.e. condition number)
and the high Pauli string count in the LCU
are expected to be relevant to QSVT based CFD algorithms.

%
\section{Acknowledgements}
The permission of Rolls-Royce to publish this paper is gratefully acknowledged.
The results were completed as part of funding received under the UK's
Commercialising Quantum Technologies Programme (Grant reference 10004857).
This work has benefited from several technical discussions and the author would like 
to thank Neil Gillespie, Joan Camps and Christoph Sunderhauf of Riverlane;
Jarrett Smalley of Rolls-Royce; and, Philippa Rubin of the STFC Hartree Centre.

%
\newpage
\bibliographystyle{ieeetr}
\bibliography{references}

\begin{thebibliography}{10}

\bibitem{walport2018computational}
M.~Walport, M.~Calder, C.~Craig, D.~Culley, R.~de~Cani, C.~Donnelly,
  R.~Douglas, B.~Edmonds, J.~Gascoigne, N.~Gilbert, {\em et~al.}, {\em
  Computational Modelling: Technological Futures}.
\newblock Government Office for Science, 2018.

\bibitem{givi2020quantum}
P.~Givi, A.~J. Daley, D.~Mavriplis, and M.~Malik, ``Quantum speedup for
  aeroscience and engineering,'' {\em AIAA Journal}, vol.~58, no.~8,
  pp.~3715--3727, 2020.

\bibitem{steijl2019quantum}
R.~Steijl, ``Quantum algorithms for fluid simulations,'' {\em Advances in
  Quantum Communication and Information}, p.~31, 2019.

\bibitem{kacewicz2006almost}
B.~Kacewicz, ``Almost optimal solution of initial-value problems by randomized
  and quantum algorithms,'' {\em Journal of Complexity}, vol.~22, no.~5,
  pp.~676--690, 2006.

\bibitem{costa2019quantum}
P.~C. Costa, S.~Jordan, and A.~Ostrander, ``Quantum algorithm for simulating
  the wave equation,'' {\em Physical Review A}, vol.~99, no.~1, p.~012323,
  2019.

\bibitem{gaitan2020finding}
F.~Gaitan, ``Finding flows of a navier--stokes fluid through quantum
  computing,'' {\em npj Quantum Information}, vol.~6, no.~1, pp.~1--6, 2020.

\bibitem{suau2021practical}
A.~Suau, G.~Staffelbach, and H.~Calandra, ``Practical quantum computing:
  solving the wave equation using a quantum approach,'' {\em ACM Transactions
  on Quantum Computing}, vol.~2, no.~1, pp.~1--35, 2021.

\bibitem{ljubomir2022quantum}
B.~Ljubomir, ``Quantum algorithm for the navier--stokes equations by using the
  streamfunction-vorticity formulation and the lattice boltzmann method,'' {\em
  International Journal of Quantum Information}, p.~2150039, 2022.

\bibitem{lu2020quantum}
C.~Lu, Z.~Hu, B.~Xie, and N.~Zhang, ``Quantum cfd simulations for heat transfer
  applications,'' in {\em ASME International Mechanical Engineering Congress
  and Exposition}, vol.~84584, p.~V010T10A050, American Society of Mechanical
  Engineers, 2020.

\bibitem{lubasch2020variational}
M.~Lubasch, J.~Joo, P.~Moinier, M.~Kiffner, and D.~Jaksch, ``Variational
  quantum algorithms for nonlinear problems,'' {\em Physical Review A},
  vol.~101, no.~1, p.~010301, 2020.

\bibitem{kyriienko2021solving}
O.~Kyriienko, A.~E. Paine, and V.~E. Elfving, ``Solving nonlinear differential
  equations with differentiable quantum circuits,'' {\em Physical Review A},
  vol.~103, no.~5, p.~052416, 2021.

\bibitem{vazquez2020enhancing}
A.~C. Vazquez, R.~Hiptmair, and S.~Woerner, ``Enhancing the quantum linear
  systems algorithm using richardson extrapolation,'' {\em arXiv preprint
  arXiv:2009.04484}, 2020.

\bibitem{lapworth2022hybrid}
L.~Lapworth, ``A hybrid quantum-classical cfd methodology with benchmark hhl
  solutions,'' {\em arXiv preprint arXiv:2206.00419}, 2022.

\bibitem{patankar1972calculation}
S.~Patankar and D.~Spalding, ``A calculation procedure for heat, mass and
  momentum transfer in three-dimensional parabolic flows,'' {\em Int. J. Heat
  Mass Transfer}, vol.~15, pp.~1787--1806, 1972.

\bibitem{ammour2018subgrid}
D.~Ammour and G.~J. Page, ``The subgrid-scale approach for modeling impingement
  cooling flow in the combustor pedestal tile,'' {\em Journal of Heat
  Transfer}, vol.~140, no.~4, 2018.

\bibitem{harrow2009quantum}
A.~W. Harrow, A.~Hassidim, and S.~Lloyd, ``Quantum algorithm for linear systems
  of equations,'' {\em Physical review letters}, vol.~103, no.~15, p.~150502,
  2009.

\bibitem{misev2018development}
C.~Misev, {\em Development and Optimisation of an Implicit CFD Solver in
  Hydra}.
\newblock University of Surrey (United Kingdom), 2018.

\bibitem{lapworth2004hydra}
L.~Lapworth, ``Hydra-cfd: a framework for collaborative cfd development,'' in
  {\em International conference on scientific and engineering computation
  (IC-SEC)}, vol.~30, 2004.

\bibitem{amirante2021multifidelity}
D.~Amirante, P.~Adami, and N.~J. Hills, ``A multifidelity aero-thermal design
  approach for secondary air systems,'' {\em Journal of Engineering for Gas
  Turbines and Power}, vol.~143, no.~3, 2021.

\bibitem{moinier2002edge}
P.~Moinier, J.-D. Muller, and M.~B. Giles, ``Edge-based multigrid and
  preconditioning for hybrid grids,'' {\em AIAA journal}, vol.~40, no.~10,
  pp.~1954--1960, 2002.

\bibitem{wesseling1982robust}
P.~Wesseling, ``A robust and efficient multigrid method,'' in {\em Multigrid
  Methods}, pp.~614--630, Springer, 1982.

\bibitem{hestenes1952methods}
M.~R. Hestenes and E.~Stiefel, ``Methods of conjugate gradients for solving,''
  {\em Journal of research of the National Bureau of Standards}, vol.~49,
  no.~6, p.~409, 1952.

\bibitem{fletcher1976conjugate}
R.~Fletcher, ``Conjugate gradient methods for indefinite systems,'' in {\em
  Numerical analysis}, pp.~73--89, Springer, 1976.

\bibitem{van1992bi}
H.~A. Van~der Vorst, ``Bi-cgstab: A fast and smoothly converging variant of
  bi-cg for the solution of nonsymmetric linear systems,'' {\em SIAM Journal on
  scientific and Statistical Computing}, vol.~13, no.~2, pp.~631--644, 1992.

\bibitem{idomura2018communication}
Y.~Idomura, T.~Ina, S.~Yamashita, N.~Onodera, S.~Yamada, and T.~Imamura,
  ``Communication avoiding multigrid preconditioned conjugate gradient method
  for extreme scale multiphase cfd simulations,'' in {\em 2018 IEEE/ACM 9th
  Workshop on Latest Advances in Scalable Algorithms for Large-Scale Systems
  (scalA)}, pp.~17--24, IEEE, 2018.

\bibitem{saad1986gmres}
Y.~Saad and M.~H. Schultz, ``Gmres: A generalized minimal residual algorithm
  for solving nonsymmetric linear systems,'' {\em SIAM Journal on scientific
  and statistical computing}, vol.~7, no.~3, pp.~856--869, 1986.

\bibitem{balay1998petsc}
S.~Balay, W.~Gropp, L.~C. McInnes, and B.~F. Smith, ``Petsc, the portable,
  extensible toolkit for scientific computation,'' {\em Argonne National
  Laboratory}, vol.~2, no.~17, 1998.

\bibitem{martyn2021grand}
J.~M. Martyn, Z.~M. Rossi, A.~K. Tan, and I.~L. Chuang, ``Grand unification of
  quantum algorithms,'' {\em PRX Quantum}, vol.~2, no.~4, p.~040203, 2021.

\bibitem{gilyen2019quantum}
A.~Gily{\'e}n, Y.~Su, G.~H. Low, and N.~Wiebe, ``Quantum singular value
  transformation and beyond: exponential improvements for quantum matrix
  arithmetics,'' in {\em Proceedings of the 51st Annual ACM SIGACT Symposium on
  Theory of Computing}, pp.~193--204, 2019.

\bibitem{dong2021efficient}
Y.~Dong, X.~Meng, K.~B. Whaley, and L.~Lin, ``Efficient phase-factor evaluation
  in quantum signal processing,'' {\em Physical Review A}, vol.~103, no.~4,
  p.~042419, 2021.

\bibitem{gough2009gnu}
B.~Gough, {\em GNU scientific library reference manual}.
\newblock Network Theory Ltd., 2009.

\bibitem{ghia1977study}
K.~Ghia, W.~Hankey, JR, and J.~Hodge, ``Study of incompressible navier-stokes
  equations in primitive variables using implicit numerical technique,'' in
  {\em 3rd Computational Fluid Dynamics Conference}, p.~648, 1977.

\bibitem{versteeg2007introduction}
H.~K. Versteeg and W.~Malalasekera, {\em An introduction to computational fluid
  dynamics: the finite volume method}.
\newblock Pearson education, 2007.

\bibitem{mazhar2016novel}
Z.~Mazhar, ``A novel fully implicit block coupled solution strategy for the
  ultimate treatment of the velocity--pressure coupling problem in
  incompressible fluid flow,'' {\em Numerical Heat Transfer, Part B:
  Fundamentals}, vol.~69, no.~2, pp.~130--149, 2016.

\bibitem{puyero1997efficient}
A.~Puyero, D.~Zingg, A.~Puyero, and D.~Zingg, ``An efficient newton-gmres
  solver for aerodynamic computations,'' in {\em 13th Computational Fluid
  Dynamics Conference}, p.~1955, 1997.

\bibitem{blais2020lethe}
B.~Blais, L.~Barbeau, V.~Bibeau, S.~Gauvin, T.~El~Geitani, S.~Golshan,
  R.~Kamble, G.~Mirakhori, and J.~Chaouki, ``Lethe: An open-source parallel
  high-order adaptative cfd solver for incompressible flows,'' {\em SoftwareX},
  vol.~12, p.~100579, 2020.

\bibitem{golub2013matrix}
G.~H. Golub and C.~F. Van~Loan, {\em Matrix computations}.
\newblock JHU press, 2013.

\bibitem{smith1975comparative}
R.~E. Smith~Jr and A.~Kidd, ``Comparative study of two numerical techniques for
  the solution of viscous flow in a driven cavity,'' {\em NASA Special
  Publication}, vol.~378, p.~61, 1975.

\bibitem{ghia1982high}
U.~Ghia, K.~N. Ghia, and C.~Shin, ``High-re solutions for incompressible flow
  using the navier-stokes equations and a multigrid method,'' {\em Journal of
  computational physics}, vol.~48, no.~3, pp.~387--411, 1982.

\bibitem{redal2019dynamfluid}
H.~Redal, J.~Carpio, P.~A. Garc{\'\i}a-Salaberri, and M.~Vera, ``Dynamfluid:
  Development and validation of a new gui-based cfd tool for the analysis of
  incompressible non-isothermal flows,'' {\em Processes}, vol.~7, no.~11,
  p.~777, 2019.

\bibitem{bouffanais2007large}
R.~Bouffanais, M.~O. Deville, and E.~Leriche, ``Large-eddy simulation of the
  flow in a lid-driven cubical cavity,'' {\em Physics of Fluids}, vol.~19,
  no.~5, p.~055108, 2007.

\bibitem{shetty2010high}
D.~A. Shetty, T.~C. Fisher, A.~R. Chunekar, and S.~H. Frankel, ``High-order
  incompressible large-eddy simulation of fully inhomogeneous turbulent
  flows,'' {\em Journal of computational physics}, vol.~229, no.~23,
  pp.~8802--8822, 2010.

\bibitem{courbebaisse2011time}
G.~Courbebaisse, R.~Bouffanais, L.~Navarro, E.~Leriche, and M.~Deville,
  ``Time-scale joint representation of dns and les numerical data,'' {\em
  Computers \& fluids}, vol.~43, no.~1, pp.~38--45, 2011.

\bibitem{lien2000pressure}
F.-S. Lien, ``A pressure-based unstructured grid method for all-speed flows,''
  {\em International journal for numerical methods in fluids}, vol.~33, no.~3,
  pp.~355--374, 2000.

\bibitem{chandra2001parallel}
R.~Chandra, L.~Dagum, D.~Kohr, R.~Menon, D.~Maydan, and J.~McDonald, {\em
  Parallel programming in OpenMP}.
\newblock Morgan kaufmann, 2001.

\bibitem{trotter1959product}
H.~F. Trotter, ``On the product of semi-groups of operators,'' {\em Proceedings
  of the American Mathematical Society}, vol.~10, no.~4, pp.~545--551, 1959.

\bibitem{suzuki1993improved}
M.~Suzuki, ``Improved trotter-like formula,'' {\em Physics Letters A},
  vol.~180, no.~3, pp.~232--234, 1993.

\bibitem{Low2019hamiltonian}
G.~H. Low and I.~L. Chuang, ``Hamiltonian {S}imulation by {Q}ubitization,''
  {\em {Quantum}}, vol.~3, p.~163, July 2019.

\bibitem{childs2012hamiltonian}
A.~M. Childs and N.~Wiebe, ``Hamiltonian simulation using linear combinations
  of unitary operations,'' {\em arXiv preprint arXiv:1202.5822}, 2012.

\bibitem{kothari2014efficient}
R.~Kothari, {\em Efficient algorithms in quantum query complexity}.
\newblock PhD thesis, University of Waterloo, 2014.

\bibitem{berry2015simulating}
D.~W. Berry, A.~M. Childs, R.~Cleve, R.~Kothari, and R.~D. Somma, ``Simulating
  hamiltonian dynamics with a truncated taylor series,'' {\em Physical review
  letters}, vol.~114, no.~9, p.~090502, 2015.

\bibitem{berry2018improved}
D.~W. Berry, M.~Kieferov{\'a}, A.~Scherer, Y.~R. Sanders, G.~H. Low, N.~Wiebe,
  C.~Gidney, and R.~Babbush, ``Improved techniques for preparing eigenstates of
  fermionic hamiltonians,'' {\em npj Quantum Information}, vol.~4, no.~1,
  pp.~1--7, 2018.

\bibitem{babbush2018encoding}
R.~Babbush, C.~Gidney, D.~W. Berry, N.~Wiebe, J.~McClean, A.~Paler, A.~Fowler,
  and H.~Neven, ``Encoding electronic spectra in quantum circuits with linear t
  complexity,'' {\em Physical Review X}, vol.~8, no.~4, p.~041015, 2018.

\end{thebibliography}

\end{document}